# Additional Acceleration and Collimation of Relativistic Electron Beams by Magnetic Field Resonance at Very High Intensity Laser Interaction

Hong Liu[1,4], X. T. He[2,3], and Heinrich Hora[5],

[1]*Graduate School, China Academy of Engineering Physics, Beijing P. O. Box 2101, Beijing 100088, P.R.China*
[2]*Institute of Applied Physics and Computational Mathematics, Beijing P. O. Box 8009, Beijing 100088, P.R.China*
[3]*Department of Physics, Zhejiang University, Hangzhou 310027, China*
[4]*[4]Basic Department, Beijing Materials Institute, Beijing 101149, China*
[5]*Department of Theoretical Physics, University of New South Wales, Sydney 2052, Australia*

In addition to the ponderomotive acceleration of highly relativistic electrons at interaction of very short and very intense laser pulses, a further acceleration is derived from the interaction of these electron beams with the spontaneous magnetic fields of about 100 MG. This additional acceleration is the result of a laser-magnetic resonance acceleration (LMRA)[1] around the peak of the azimuthal magnetic field. This causes the electrons to gain energy within a laser period. Using a Gaussian laser pulse, the LMRA acceleration of the electrons depends on the laser polarization. Since this is in the resonance regime, the strong magnetic fields affect the electron acceleration considerably. The mechanism results in good collimated high energetic electrons propagating along the center axis of the laser beam as has been observed by experiments and is reproduced by our numerical simulations.

Key Words: Electron acceleration, Intense laser, Strong quasistatic magnetic field, Laser-plasma interaction

PACS numbers: 41.75.Jv 52.38.Kd 52.65.Cc

## 1. INTRODUCTION

The use of very short laser pulses of picosecond or less duration with intensities of TW and up of PW and beyond arrived of a new category of interactions. The phenomena to be discussed now are polarization dependence effects[1,2,3,4,5,6,7,8,9,10], deviations of the generated plasmas from nearly space charge neutralization as in the earlier cases, and in relativistic effects. The earlier observed cases could mostly be described by space neutralized plasma hydrodynamics even including relativistic self focusing[11,12,13,14,15] generated ions of several 100 MeV energy because the Debye lengths involved were sufficiently short and the internal electric fields[16] were not yet of dominating influence. The acceleration of free electrons by laser fields was well discussed separately [15,17,18] and resulted in some agreement with ps-TW measurements[19]. Nevertheless very high density relativistic electron beams were measured recently which resulted in a new situation where relativistic self-focusing, plasma motion, and the beam generation described by particle-in-cell (PIC) methods [10] were covering the phenomena not completely depending on each situation.

The generation of extremely high magnetic fields in laser-produced plasmas [20] was known since a long time but there is a basically new situation with the very intense relativistic electron beams and their mutual interaction with the very high magnetic fields. We present here studies of these interactions of the electron-beams with the magnetic fields as a basic new laser-magnetic resonance acceleration (LMRA) mechanism which results in a kind of pinch effect with a high degree of collimation of the electron beams. As the fast-ignitor (FI) concept[21] for inertial confinement fusion relies on many new phenomena such as explosive channel formation[22] and self-generated huge magnetic field[20], much interest in ultraintense laser-plasma interaction studies appeared. More interestingly, the gigagause self-generated azimuthal magnetic field has been observed recently[9].

We found that two different fast electrons exist in the presence of self-generated azimuthal magnetic field. Two sources of multi-MeV electrons are distinguished from the relativistic laser-plasma interaction. The first is that the electron acceleration depends on the laser intensity, known as the pondermotive acceleration [17,16,18,19]. The second is that, around the peak of azimuthal magnetic field, LMRA partly occurs which causes the electron to gain energy from the ratio between electron Larmor frequency and laser frequency within one laser period[23]. If we consider a linearly polarized (LP) laser pulse, the LMRA results in a dependence of the laser accelerated electrons on the laser polarization. Because in the resonance regime, the strong magnetic field affects the electron acceleration dramatically. Only just from the second source, polarization dependence of electron is appeared. In our knowledge, these different sources of fast electrons are mentioned for the first time. This clears up many experiments and PIC simulations which are related with polarization dependence phenomena.

A fully relativistic single particle code is developed to investigate the dynamical properties of the energetic electrons. The single test electron model is a simple but effective one. It has been used to analyze the direct laser acceleration of relativistic electrons in plasma channels, e. g. M.Schmitz *et al.*[24] have analyzed the LP laser system with self-generated static electric field and discussed the electron resonant acceleration mechanism. K.P.Singh[25] found that resonance occurs between the electrons and electric field of the laser pulse. We find a big difference between the pondermotive acceleration and LMRA mechanism in this paper. We discuss LMRA mechanism of electrons in strong LP laser pulse and self-generated azimuthal magnetic field. In our simulation, the laser field is a Gaussian profile and the quasistatic magnetic field is in a circle profile which lies on the laser intensity and as a function of circle radius. Our paper is organized as follows. In Sec.II, we describe the dynamical behavior of relativistic electrons in a combined LP laser and quasistatic azimuthal magnetic field numerically. Two regimes (the peak of laser and the peak of quasistatic magnetic field) and two typical directions (in polarization direction and $90^0$ turned from the polarization direction) are discussed. We also show an approximate analytical equation which provides a full understanding of the LMRA mechanism for comparison. The numerical results clearly demonstrate that LMRA partly occurs within one laser period. Our discussion and conclusion are given in Sec.III.

## II. GAUSSIAN LP LASER PULSE MODEL

The relativistic Lorentz force equations with quasistatic magnetic-field which is perpendicular to the laser propagation direction are

$$\frac{d\mathbf{p}}{dt} = \frac{\partial \mathbf{a}}{\partial t} - \mathbf{v} \times (\nabla \times \mathbf{a} + \mathbf{b_\theta}), \qquad (1)$$

$$\frac{d\gamma}{dt} = \mathbf{v} \cdot \frac{\partial \mathbf{a}}{\partial t}, \qquad (2)$$

where $\mathbf{a}$ is the normalized vector potential. $\mathbf{b}_\theta$ is the normalized azimuthal magnetic field, $\mathbf{v}$ is the normalized velocity of electron, $\mathbf{p}$ is the normalized relativistic momentum, $\gamma = (1-v^2)^{-1/2}$ is the relativistic factor or normalized energy. Their dimensionless forms are $\mathbf{a} = \frac{e\mathbf{A}}{m_e c^2}$, $\mathbf{b} = \frac{e\mathbf{B}}{m_e c \omega}$, $\mathbf{v} = \frac{\mathbf{u}}{c}$, $\mathbf{p} = \frac{\mathbf{P}}{m_e c} = \gamma \mathbf{v}$, $t = \omega t$, $r = kr$, $m_e$ and $e$ are the electric mass and charge, respectively, $c$ is the light velocity. $k = 2\pi/\lambda$ is the wave number, $\lambda$ is the wave length. We assume that the laser propagation is in positive $\hat{\mathbf{z}}$ direction along the plasma channel with a phase velocity $v_{ph}$. For simplicity, in the following discussions we assume that the phase velocity of the laser pulse equals to the light velocity, i.e. $v_{ph} = c$. The main results obtained can be readily extended to the case of $v_{ph} \neq c$. For the irradiance of the femtosecond laser pulses, the plasma ions have no time to respond to the laser and therefore can be assumed to be immobile. We have used the Coulomb gauge. Here, because the dimensionless self-generated azimuthal magnetic field e.g. 2 is much larger than the static electric field e.g. 0.01, the effects of the static electric field can be ignored.

For a linearly focused Gaussian profile laser with frequency $\omega$ along in plasma channel can be modeled as

$$\mathbf{a} = a_0 e^{-\frac{x^2+y^2}{R_0^2}} \cdot e^{-\frac{(kz-\omega t)^2}{k^2 L^2}} \cdot \cos(kz - \omega t)\hat{\mathbf{x}} \qquad (3)$$

where the critical density $n_c = m\omega^2/(4\pi e^2)$, the plasma frequency equals to the light frequency, $L$ and $R_0$ are the pulse width and minimum spot size, respectively. Laser pulse is a transverse wave satisfying $\mathbf{k} \cdot \mathbf{a} = 0$. Its profile shows in Fig.1(a).

For the background field, the generation of azimuthal quassistatic magnetic field ($\mathbf{b}_\theta$) has been discussed by many authors and observed in experiments. The typical work related with ultraintense short laser pulse with overdense plasma interaction has been done by R. N. Sudan[20] in 1993. He proposed that the mechanism for magnetic field generation is a result of dc currents driven by the spatial gradients and temporal variations of the ponderomative force exerted by the laser on the plasma electrons. In recent experiment, M. Tatarkis *et al.* observed the peak of $\mathbf{b}_\theta$ at least hundreds of MG as given in Ref. 9. Our model uses $\mathbf{b}_\theta$ in the form:

$$\mathbf{b}_\theta = -b_{\theta 0} \cdot r \cdot <a^2> \hat{\boldsymbol{\theta}} \qquad (4)$$

The above explicit expression clearly indicates that the self-generated magnetic field $\mathbf{b}_\theta$ is an oriented circle. It caused by the longitudinal electron currents motion. We have assumed that ion immobile. $b_{\theta 0}$ is an approximately coefficient, including slow-time changed plasma parameters. $r = \sqrt{x^2 + y^2}$ is the distance from the axis. $<\ >$ denotes time average over one laser period. $r \cdot <a^2>$ decide the structure of $\mathbf{b}_\theta$. The peak of $\mathbf{b}_\theta$ is located at the $R_0/2$ laser spot. Although the reality is far more complex and the form will be significantly different. We use the rough profile to investigate the dynamics of the fast electron. Its profile shows in Fig.1(b).

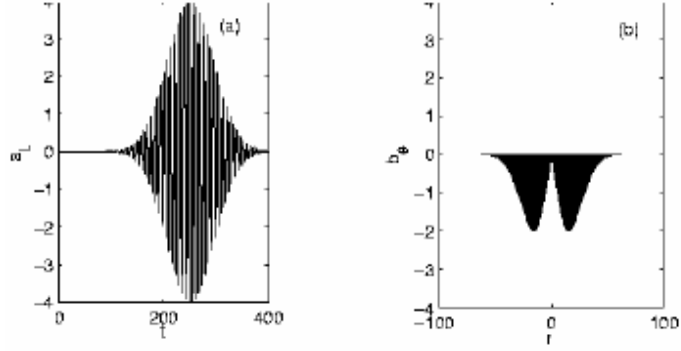

Fig. 1 (a) The profile of laser intensity $a$ as a function of time in the units of $\omega^{-1}$, (b) The profile of quasistatic magnetic field $b_\theta$ (units 1 corresponding to $100MG$) as a function of circle radius in the units of $k^{-1}$.

We assume that the trajectory of a test electron starts at $\mathbf{v}_0 = 0$. Eq.(1) and Eq.(2) yield

$$\frac{dp_x}{dt} = (v_z - 1)\frac{\partial a}{\partial z} + v_y \frac{\partial a}{\partial y} + v_z b_\theta \cos\theta \tag{5}$$

$$\frac{dp_y}{dt} = -v_x \frac{\partial a}{\partial y} + v_z b_\theta \sin\theta \tag{6}$$

$$\frac{dp_z}{dt} = -v_x \frac{\partial a}{\partial z} - v_x b_\theta \cos\theta - v_y b_\theta \sin\theta \tag{7}$$

$$\frac{d\gamma}{dt} = -v_x \frac{\partial a}{\partial z} \tag{8}$$

where $\mathbf{b}_\theta = b_\theta(-\sin\theta\hat{\mathbf{x}} + \cos\theta\hat{\mathbf{y}})$, $\theta = arctg(\frac{y}{x})$. An exact analytical solution of Eqs.(5)-(8) is impossible because of their nonlinearity. Nevertheless, these equations reveal the mechanism of acceleration and collimation and will be solved numerically.

Using Eqs.(5)-(8), we choose different initial position to investigate the electron dynamics of a LP Gaussian profile laser pulse. Because the initial velocity can be transformed to initial position in our single test electron case, we keep initial velocity at rest and change the initial positions of the test electrons. We assume that the trajectory of a test electron starts from $\mathbf{v}_0 = 0$ and $z_0 = 4L$ at $t = 0$, while the center of laser pulse locates at $z = 0$, then the classical trajectory is then fully determined by Eqs.(5)-(8). Now we choose following parameters that are available in present experiments, i.e. $L = 10\lambda$, $R_0 = 5\lambda$ ($\lambda = 1.06\mu m$), $a_0 = 4$ (corresponding to $I \approx 2\times10^{19}W/cm^2$), $b_{\theta 0} = 2$ (corresponding to $B_\theta \approx 200MG$). Then, we trace the temporal evolution of electron energy and trajectory and plot the results in Fig.2-3.

In order to explain the simulation results, we excerpt an analytical equation which has been obtained in Ref. [23]

$$\gamma = 1 + \frac{1}{2}\frac{a^2}{(1+\frac{b}{\omega})^2}. \tag{9}$$

where $a$ and $b$ are the local magnitude of the laser and the quasistatic magnetic field of the energetic electron. Although the equation is derived from a model which contains a circularly polarized laser and an axial static magnetic field, it indicates the resonance between the laser field and the magnetic field. Because the energy Eq.(9) is independence on time, we use it to explain what drives the energy of electron to high energy along the strong magnetic field presence. That is when the LMRA occurs or partly occurs within laser period, the electron will gain energy from the near resonance point (singularity) at a negative $b_\theta (\approx \omega)$. The electron acceleration depends not only on the laser intensity, but also on the ratio between electron Larmor frequency and laser frequency.

## III. NUMERICAL RESULTS

Fig.2 and Fig.3 show the track of a test electron and its correspondent net energy gain in the combined $\mathbf{a}$ and $\mathbf{b}_\theta$ fields from different regimes, e.g. the peak of laser and the peak of quasistatic magnetic field respectively. (a)-(c) The trajectory of electron start at $x_0 = \lambda$, $y_0 = 0$ (in polarization direction $\hat{\mathbf{x}}$), (b)-(d) The trajectory of electron start at $x_0 = 0$, $y_0 = \lambda$ ($90^0$ turned from the direction of polarization). The difference trajectory of the test electron and its correspondent energy gain from the different initial positions can be compared. We also show the electron energy $\gamma$ (in the units of $mc^2$) as a function of time (in the units of $\omega^{-1}$) in dashed line for the case of without $\mathbf{b}_\theta$ in Fig.2-3(c) and (d). We like to emphasize that at relativistic intensities laser ($I > 10^{18} W/cm^2$) the electron drift velocity is very slow but not slow enough. In fact, for $a_0 = 2$ the drift velocity is the same order of the quiver velocity. When $a_0 = 4$, $\mathbf{p}_\perp = \mathbf{a}$, $p_z = \frac{a^2}{2}$ from $\gamma = \sqrt{1 + p_\perp^2 + p_z^2}$, then $\gamma_{max} = 9$. The nonlinear ponderomotive scattering angle[19] (in vacuum) $\theta = arctg\sqrt{\frac{2}{\gamma-1}} \approx 28^0$. The electron momentums of two transverse directions are independent on the laser polarization. But such large scattering angles will be unfavorable to the fast ignition of the high compressed fuel. When a strong self-generated azimuthal magnetic field presence, things will be changed. When $\mathbf{a}$ and $\mathbf{b}_\theta$ coexist, e.g. $a = 4$, $b_\theta = -0.6$, using our derived Eq.(9) we can estimate that $\gamma_{max} \approx 51$ which is very close to our simulation results. Within one laser period, the LMRA mechanism can partly occur and give chance to let electron rest in one phase of the laser for a while. This relatively rest makes the electron in a slowly motion and gain energy from the laser field. The efficiency of energy transfer will be high. This is the laser-magnetic resonance acceleration. It's very different from the pondermotive acceleration which does not concern the laser period. In our profile of $\mathbf{b}_\theta$, a lower component quassistatic magnetic field exists in the center regime. So the electron energy gain is high than the ponderomotive acceleration energy. Anyway in this regime ($r \leq \lambda$) ponderomotive acceleration is in dominant and polarization independence still remain.

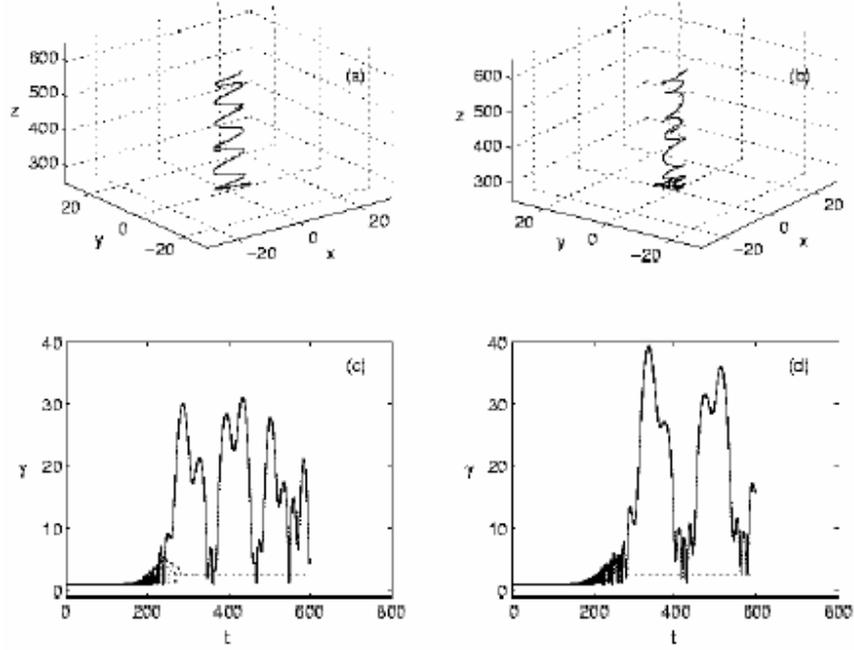

Fig. 2 Electron in combined **a** and $\mathbf{b}_\theta$ fields. (a) The trajectory of electron start at $x_0 = \lambda$, $y_0 = 0$ (in polarization direction $\hat{\mathbf{x}}$), (b) The trajectory of electron start at $x_0 = 0$, $y_0 = \lambda$ ($90^0$ turned from the direction of polarization). (c) and (d) Electron energy $\gamma$ in units of $mc^2$ as a function of time in the units of $\omega^{-1}$. Other parameters is corresponding to (a) and (b) respectively, with $\mathbf{b}_\theta$ (solid line) and without $\mathbf{b}_\theta$ (dashed line).

Fig.3 the parameters of initial position were changed to (a) $x_0 = R_0/2$, $y_0 = 0$, (b) $x_0 = 0$, $y_0 = R_0/2$ (around the peak of $\mathbf{b}_\theta$). One can find that evidence deform appears in solid line between Fig.3(a)-(c) and (b)-(d). Polarization dependence is a main feature in this regime. If the initial position is in the polarization direction, the electron has quiver energy to let the LMRA occurs, otherwise when the initial position is not in polarization direction, the electron has no quiver energy to utilize. This is the one reason which makes the low efficiency of energy transfer than circularly polarized (CP) laser case. When **a** and $\mathbf{b}_\theta$ coexist, e.g. $a = 2$, $b_\theta = -0.85$, using our derived Eq.(9) we can estimate that $\gamma_{max} \approx 89$ which is very close to our simulation results. Evident, pondermotive acceleration still be shown in dot line in Fig.3(c) and (d), its value relatively smaller than that in the center regime. The electron polarization dependence is controlled by the competition of the amplitude of **a** and $\mathbf{b}_\theta$. If the value of **a** is in dominant, e.g. in center regime, the polarization dependence is not evident, shows in Fig.2. If the value of $\mathbf{b}_\theta$ is large than **a**, e.g. in the second regime, polarization dependent appears which shows in Fig.3. Because of the different initial position in or not in polarization direction, the electron has different chance to utilize quiver energy and make LMRA to occur.

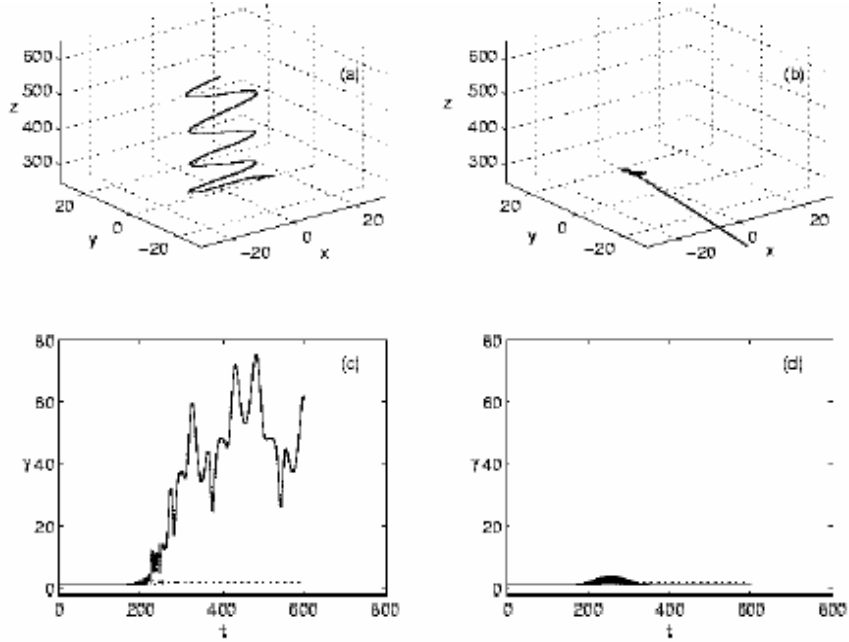

Fig. 3 The parameters are same with Fig.2 but only (a) $x_0 = R_0/2$, $y_0 = 0$ (in the peak of $\mathbf{b}_\theta$) (b) $x_0 = 0$, $y_0 = R_0/2$ ($90^0$ turned from the direction of polarization).

Our simulations satisfy with the phenomena which have been reported by experiments and numerical simulations e.g. [4,5,6,7,10]. For example in Ref.[4] authors pointed out that a narrow plasma jet is formed at the rear surface which is consistent with a beam of fast electrons traveling through the target, collimated by a magnetic field in the target. In Ref.[5] L.Gremillet et al. observed two narrow long jets originating from the focal spot. These may be caused by the $\mathbf{b}_\theta$ in the second region. Even the $\mathbf{b}_\theta$ has more than one peak, more electron jets can be produced. As given in Ref.6 the snake like electron orbit is very similar to our Fig.2(a), (b) and Fig. 3(a). If the amplitude of $\mathbf{a}$ and $\mathbf{b}_\theta$ can be comparable, an elliptical heating area appears, such as pointed out by Kodama et al. in experiment [7]. In Ref.[10] A. Pukhov et al. pointed out that distribution of electron current $\mathbf{J}$ and quasistatic magnetic field $\mathbf{B}$ at the positions of tight focusing is elongated in the direction of polarization and heavy relativistic electrons sprayed in the direction of polarization. From above analysis, the polarization dependence of LP system is a typical different feature with a circularly polarized (CP) system when the self-generated azimuthal magnetic fields are present..

### IV. DISCUSSIONS AND CONCLUSIONS

Using a single test electron model, we study the energetic electrons in combined strong azimuthal magnetic field and Gaussian profile linearly polarized laser field numerically. Two different source of fast electron are distinguished. In the presence of magnetic field in LP system, polarization independence is being modified by the increasing value of magnetic field. If the laser intense is in dominant, the polarization dependence is not evident, If the value of magnetic field becomes comparable with the laser intensity, the polarization dependent appears. Comparing with an energy analytic solution of electron which contains the laser-magnetic resonance acceleration mechanism, we point out that strong quasistatic magnetic field affect electron acceleration dramatically from the ratio between the Larmor frequency and the laser frequency. As the ratio approaches unity, clear resonance peaks are observed. From the physical parameters available for laboratory experiments, we find that the electron acceleration depends not only on the laser intensity, but also on the ratio between electron Larmor frequency and the laser frequency. The different fast electrons which produce by LMRA and pondermotive acceleration mechanism give an clear explain of the polarization dependent phenomena. Because the LMRA relates with laser period, an averaged calculation over one laser period will lost the effect of $\mathbf{b}_\theta$. This is different from the pondermotive acceleration mechanism.

For the study of relativistic strong laser pulse along with a hundreds of MG azimuthal quassistatic magnetic field is a complex process, related with several mechanisms. In this paper we simply treat the laser pulse in channel and the quassistatic magnetic fields, even not consider the energetic electrons interact with and are deflected by background particles. Our purpose is to make clear how the fast electron behavior in the presence of a magnetic field. For the polarization dependent phenomena only appears in LP laser case, whereas the CP laser case hasn't, so the efficiency of energy transfer will be different. The value in CP case is higher than in LP case. For the quassistatic magnetic field modifying the polarization independence is very important in laser-plasma interactions, it will have good application in fast-ignitor scheme and particle accelerators.

## V. ACKNOWLEDGMENTS

This work was supported by National Hi-Tech Inertial Confinement Fusion Committee of China, National Natural Science Foundation of China, National Basic Research Project nonlinear Science in China, and National Key Basic Research Special Foundation.